# Graphene porous foams for capacitive pressure sensing

Lekshmi A. Kurup†, Cameron M. Cole†, Joshua N. Arthur‡, and Soniya D. Yambem† *

†School of Chemistry and Physics, Centre for Materials Science, Faculty of Science, Queensland University of Technology (QUT), 2 George Street, Brisbane, QLD 4000, Australia.
E-mail: soniya.yambem@qut.edu.au

Flexible pressure sensors are an attractive area of research due to their potential applications in biomedical sensing and wearable devices. Among flexible and wearable pressure sensors, capacitive pressure sensors show significant advantages, owing to their potential low cost, ultra-low power consumption, tolerance to temperature variations, high sensitivity, and low hysteresis. In this work, we develop capacitive flexible pressure sensors using graphene based conductive foams. In these soft and porous conductive foams, graphene is present either as a coating of the pores in the foam, inside the structure of the foam itself, or a combination of both. We demonstrate that they are durable and sensitive at low pressure ranges (<10 kPa). Systematic analysis of the different pressure sensors revealed that the porous foams with graphene coated pores are the most sensitive (~ 0.137 kPa-1) in the pressure range 0 – 6 kPa. Additionally, we achieved very low limit of detection of 0.14 Pa, which is one of the lowest values reported. Further, we demonstrated the potential applications of our pressure sensors by showing detection of weak physiological signals of the body. Our work is highly relevant for research in flexible pressure sensors based on conductive foams as it shows the impact of different ways of incorporating conductive material on performance of pressure sensors.

## 1. Introduction

Flexible pressure sensors are transducers that convert mechanical deformation into electrical signals. In recent years they have been widely researched for wearable electronics applications such as electronic skin,[1-3] health care technologies,[4-5] touch screen displays,[6-7] and human-machine interactive systems.[8-9] A variety of pressure sensors with flexible and wearable characteristics based on different transducing principles have been explored including piezoresistive,[10-11] piezoelectric,[12-13] triboelectric,[14-15] and capacitive[1,9] pressure sensors. Among these, capacitive pressure sensors have the advantage of high tolerance to temperature variations, potential low cost, good dynamic response, fast response time, and high durability.[1,5,16]

For capacitive pressure sensors, which work on the simple principle of parallel plate capacitance, the flexibility and sensing property of the device depends on the mechanical property of the dielectric material.[16-18] The most common method to enhance the flexibility of capacitive pressure sensors is to use elastomeric materials with a low Young's modulus to increase deformation at lower forces and therefore, increase the dielectric response (sensitivity). Among dielectric materials, polydimethylsiloxane (PDMS) is considered the most suitable candidate due to its low cost, biocompatibility, and flexibility. However, pressure sensors with simple and unstructured PDMS films have low sensitivity.[19] Unstructured PDMS also causes slow relaxation time after the removal of pressure, which is not desirable for pressure sensors.[20] To combat these problems, there are reports of PDMS which are architecturally modified with microstructures of various geometries such as domes,[21-22] pillars,[1,23] and pores.[16,18,24] These microstructures increase the air gap between the electrodes, increasing the permittivity of the material and has been reported as one of the most effective strategies.

Another approach for increasing permittivity in PDMS microstructures is the introduction of conductive nanoparticles or high-k nanofillers such as nanoparticles of carbon and its derivatives, which resulted in pressure sensors with higher sensitivity, faster response time and improved stability.[16, 18, 25-26] Recently, graphene and its various forms as conductive filler have attracted much attention for potential applications in pressure and strain sensing, due to the ultra-sensitivity that can be achieved.[27] As such, there are numerous reports on graphene based porous piezo-resistive pressure sensors. However, there are only few reports on capacitive pressure sensors with graphene based porous foams. And in these reports of PDMS foam based capacitive pressor sensors, graphene is generally embedded in the PDMS matrix.[28] It has also been shown that conductive fillers (especially carbon black)[16, 18] which are present as a coating of the pores in the polymer foam increases sensitivity for capacitive pressure sensors. In summary, separate reports on the different ways of incorporating graphene in different materials of foams exists. But to understand the influence of the way graphene is present in porous foams on performance of capacitive pressure sensors, a comparative investigation with the same material of the foam is necessary.

In this paper, we explored different ways of integrating graphene in a PDMS soft, porous foam. Graphene is present either, (a) on the surface of the pores of the foam, (b) embedded within the foam, or a combination of both (a) and (b). The porous nature and the elasticity of PDMS, as well as the presence of graphene flakes makes the pressure sensors highly suitable for monitoring weak biomedical signals, which is demonstrated in this work. Our findings are significant and open a way for future development of sensing units that are low-cost, low power consuming and have high sensitivity, for applications in flexible biosensors.

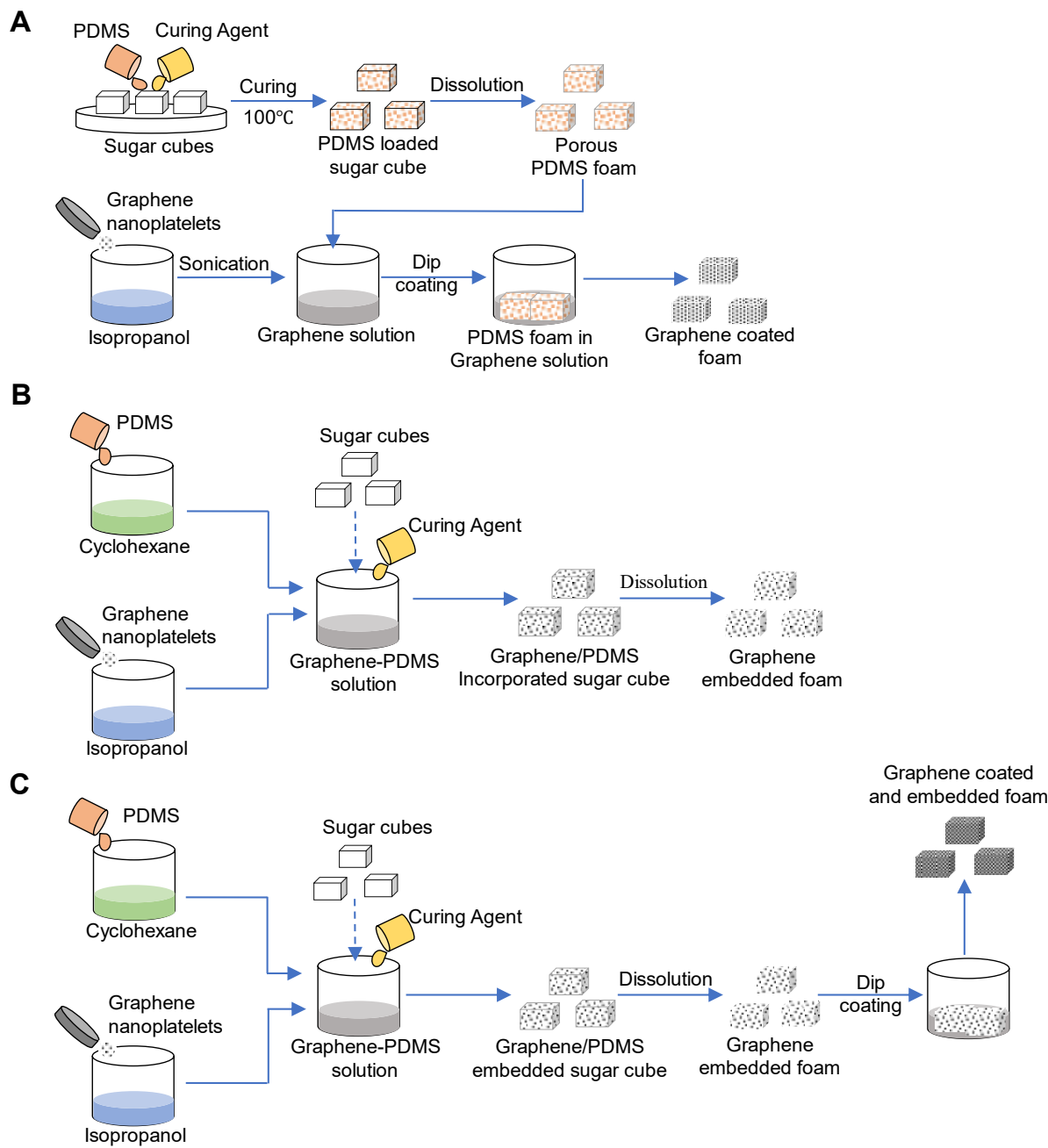

**Figure. 1** Schematics illustrating the fabrication process of **(A)** graphene coated foam (GCF) where graphene is present as coatings of the pores in the foam; **(B)** graphene embedded foam (GEF) where graphene is present in the structure of the foam; and **(C)** graphene coated and embedded foam (GCEF), where graphene is both coated on to the pores and added into the structure of the foam.

## 2. Results and discussion

### 2.1. Fabrication of graphene porous foams

To make porous foams we used a sugar templating method,[29] which is a simple and low-cost fabrication technique. Three different types of porous foams with different methods of incorporating graphene in the foam, were fabricated. In the first type, graphene is coated onto the pores of the foam. **Figure 1**A illustrates the two-step fabrication process for the graphene coated foam (GCF). Firstly, a porous PDMS foam was developed from the sugar cube templates. Secondly, the porous PDMS foam was coated with graphene by dipping it in graphene dispersed in isopropanol. The second type of foam has graphene embedded in the structure of the foam. To make this graphene embedded foam (GEF), graphene dispersed in isopropanol is mixed with a PDMS precursor, which is dissolved in cyclohexane (Figure 1B), followed by the addition of a curing agent. The sugar cube templates were then dropped in this mixture to create the GEF. The third type of foam is a combination of the first and second type. This foam, graphene coated and embedded foam (GCEF), has graphene present both as coatings of the pores in the foam as well as embedded in the structure of the foam itself. The fabrication of GCEF is also a two-step process, where a GEF was developed and dipped into a graphene solution to form the GCEF, as illustrated in Figure 1C. Photographs of the steps of fabrication for the graphene porous foams (GPFs) are shown in Figure S1-3.

## 2.2. Characteristics of graphene porous foams

Scanning electron microscopy (SEM) images showing the topography and cross-sectional views of the GCF and the GEF are provided in **Figure 2**A-D. These images (Figure 2A & C) confirm that the foams are highly porous, with the pores being largely interconnected with some exceptions. The high porosity is also evident when observed through a magnified lens (Figure S4). In Figure 2B, the GCF has a rough and irregular surface, which indicates the presence of graphene flakes attached on the pore walls of the foam. On the other hand, for GEF, as seen in Figure 2D, the surface is visibly smooth, with fewer irregularities. This is expected as graphene flakes are embedded within the PDMS microstructures. The surface of GCEF would essentially look the same as GCF foams and hence SEM image are not included.

The porosity, $P_a$, of the GPFs were determined using gravimetric measurements in air according to Equation 1.

$$P_a = \frac{V_{air\ pores}}{V_{foam\ composite}} = \frac{V_{foam\ composite} - V_{PDMS}}{V_{foam\ composite}} \qquad 1$$

where $V_{foam\ composite}$ represents the volume of the porous foam, $V_{PDMS}$ represents the volume of PDMS, and $V_{air\ pores}$ represents the volume of air pores. All porous foams have a calculated porosity ~ 80%, which indicates a highly porous and lightweight structure. Similar porosity has been reported for porous foams obtained from sugar templating.[30] Detailed porosity calculations are provided in the supporting information. The ultra-lightweight nature of the GPF is also clearly visible from the photographs in Figure 2E & F, when the foam is held on top of a dandelion flower (with seeds ready to scatter) and soft petals of a small flower (Gazania flower). As a result of the very high porosity, the foam is also extremely compressible and highly flexible, as shown by the photographs in Figure 2G & H.

**Figure 3** illustrates the average compressive stress-strain curves for the three different foams with 2 mm thickness. All stress-strain figures indicate two regions, where region I corresponds to a large increase in strain with small increase in the stress. This is where the pattens (plates of the instrument) makes full contact with the surface of the foam followed by the deformation of the cell walls of the pores by elastic bulking and gradual closing of the pores. Region I extend to till about the pores are completely closed.[25] Region II corresponds to a reduced rise in strain with large change in stress, which corresponds to the non-linear elastic deformation caused by the complete closure of pores, strain hardening and bulking of PDMS.[25] Stress-strain relationship for 4 mm thick foams are shown in Figure S5 and are very similar to that of 2 mm thick GPFs.

After the loading and unloading cycle, the thicknesses of the foams are reduced by a small amount, which is also reflected in the stress-strain curves in Figure 3. In the stress-strain curves, the strain does not reach zero when the stress is reduced to 0% in the unloading cycle. This residual strain is highest for GCF (Figure 3A) as compared to GEF and GCEF (Figures 3B and 3C, respectively). This variation in residual strain is because of the variation in the stiffness of the GPFs, which is solely due to the difference in how graphene is incorporated in the foams, as all the GPFs are developed from the same type of sugar cubes, and hence have the same porosity. GCF has the least stiffness since it has graphene only as coatings of the pores of the foam while GEF and GCEF have graphene embedded in the PDMS matrix, giving them additional mechanical strength. The difference in stiffness of the GPFs is also seen from the difference in the maximum strain %. The average maximum strain % for GCFs are highest followed by GEFs and GCEFs. Average maximum strain % for GCFs, GEFs and GCEFs are 82.7 % (71.9 %), 72.3 % (66.9 %), and 66.7 % (61.1 %) respectively, for 2 mm (4 mm) thick foams (Figure S5, Table S1). This trend in statistical data for the GPFs is expected. For GCEFs, graphene is present both inside the PDMS matrix and as coating of the pores. Therefore, it will have the maximum mechanical strength and will be stiffer, which can withstand more stress, and hence the least average strain %. The GEFs have a lower strain % than GCEFs since it has graphene only in the PDMS matrix. However, it still has a higher mechanical strength than the GCFs, which has the highest strain %. The GCFs have graphene flakes on the walls of the pores (outside the PDMS matrix) and hence lower mechanical strength compared to the other two foams.

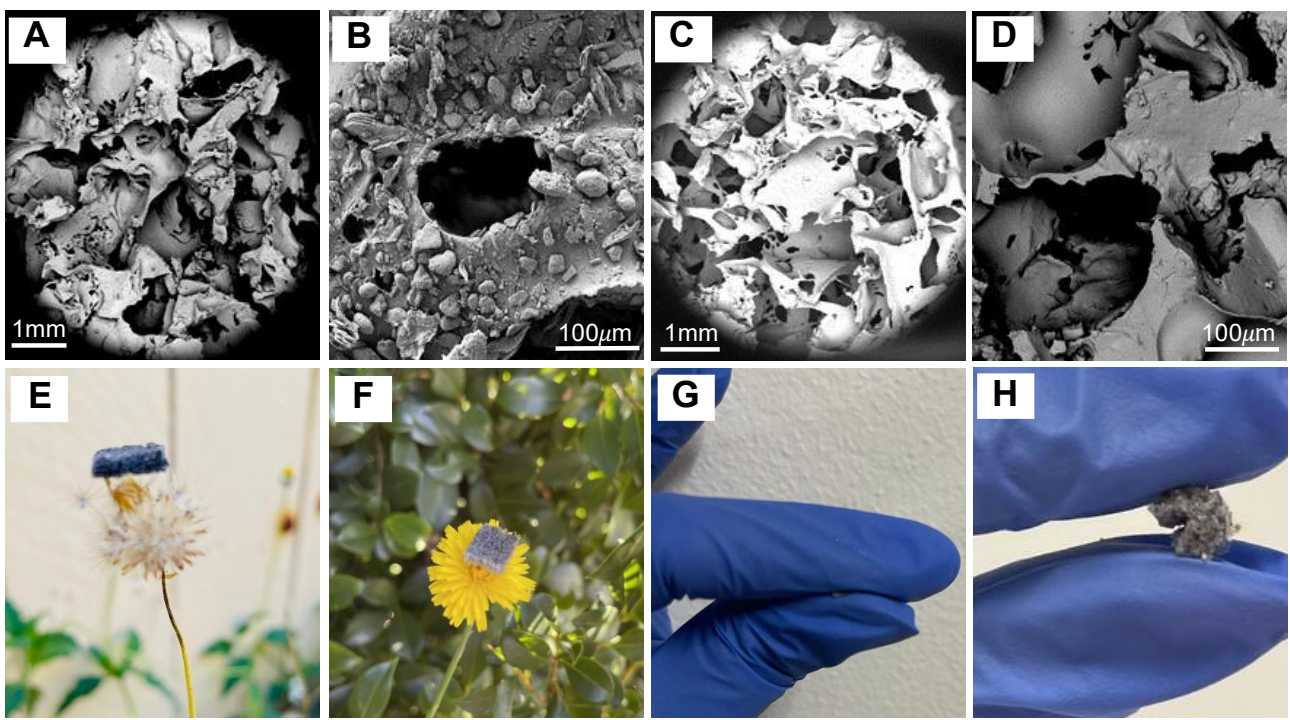

**Figure 2**. SEM images of, **(A, B)** GCF, showing the rough surface which is due to presence of graphene coated on the pores of the foam and **(C, D)** GEF, showing smooth surfaces when graphene is present in the structure of the foam. Photographs showcasing the characteristics of the foams: **(E, F)** ultra-light weight; **(G)** highly compressible; and **(H)** highly flexible.

## 2.3. Internal mechanism

**Figure 4** provides a schematic illustration of structural changes in the foam when a downward pressure is applied. The most obvious change is the closing of pores (Figures 4B & D), reducing the amount of air in the foam which leads to an increase in dielectric constant of the foam. Upon application of pressure, the orientation and/or distribution of the graphene platelets in the foam will also change. For the sake of simplicity of discussion, we assume that graphene flakes are distributed evenly in all the pores, denoted by elongated granules in Figures 4E-F. In reality, it will be a mixture of graphene flakes and clusters of flakes. Even then the explanations and reasoning are still valid. In GCFs, the graphene platelets are present outside the foam covering the inner pore walls (Figure 4E). Initially in an uncompressed state (Figure 4E), the graphene flakes are randomly aligned on the walls of the pores, with some overlapped over small contact area while some flakes stay separately attached to the walls. Gradually, with the increase in the applied pressure, the graphene flakes will come closer and in more contact with each other (Figure 4F). Under applied pressure platelets are more densely packed and the contact area between platelets in increased, resulting in a percolation network.[31,32] As a result, when incorporated in a pressure sensor, this would improve the transport of electrons as there are increased conductive pathways. On further increase in the pressure, the air pores will collapse, replacing the lower dielectric constant medium (air) with a higher dielectric constant of the solid, compressed PDMS. This collapsing of the graphene filled pores results in stress-induced tunnelling, as electrons starts tunnelling through the PDMS barrier when the distance between the graphene flakes are 2 nm or less.[33-35]

In GEFs, the graphene flakes are embedded within the PDMS matrix (Figure 4G), where some flakes are not in contact with each other, and some are already in direct contact with each other but not forming a continuous network throughout the matrix. The tunnelling effect will be present for the graphene flakes which are less than 2 nm apart,[33-35] which is the main conductive pathway for GEFs in the non-compressed state. On application of the pressure, the pores start to collapse, altering the dielectric property of the foam. At the same time, the compression reduces the thickness of the PDMS barrier between graphene flakes, which facilitates more tunneling of electrons (Figure 4H). With further increase in pressure, there will be complete compression of the pores and noticeable compression of the PDMS bulk. As a result, almost all the flakes will be at minimal distance between each other resulting in multiple conductive (tunnelling) pathways.[36]

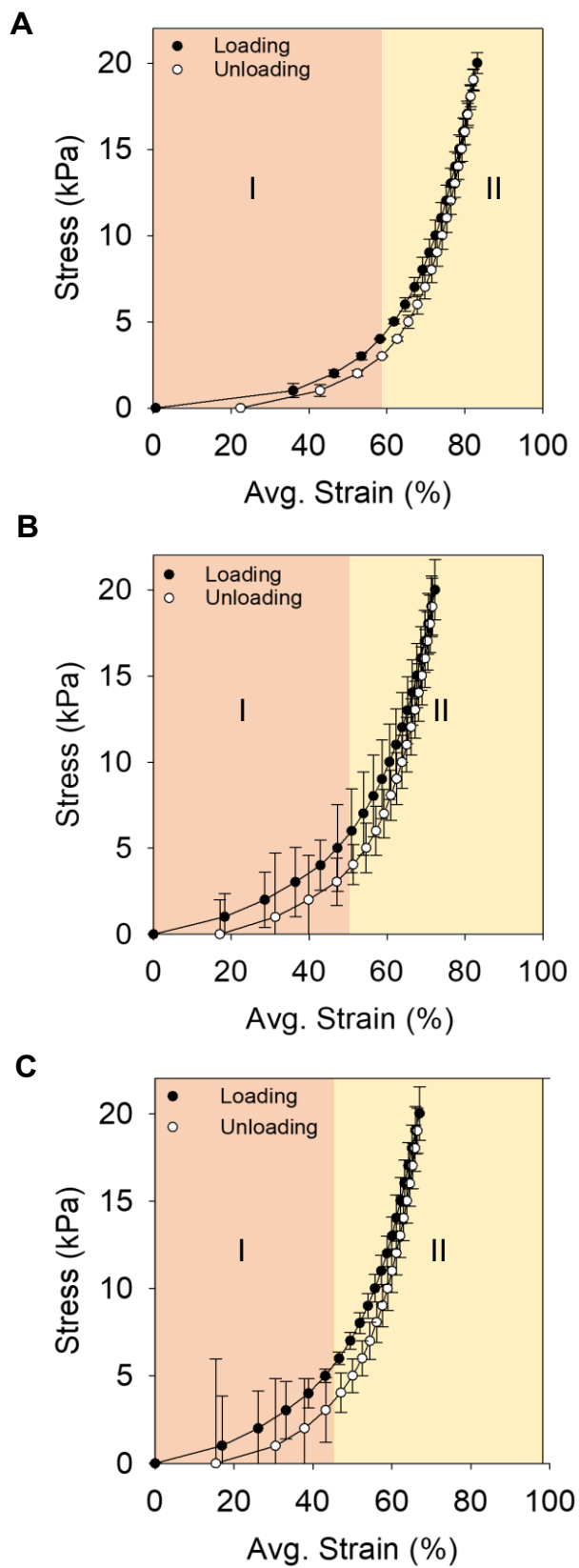

**Figure 3.** Stress-Strain curves with a maximum stress of 20 kPa for 2 mm thick **(A)** GCF; **(B)** GEF; and **(C)** GCEF.

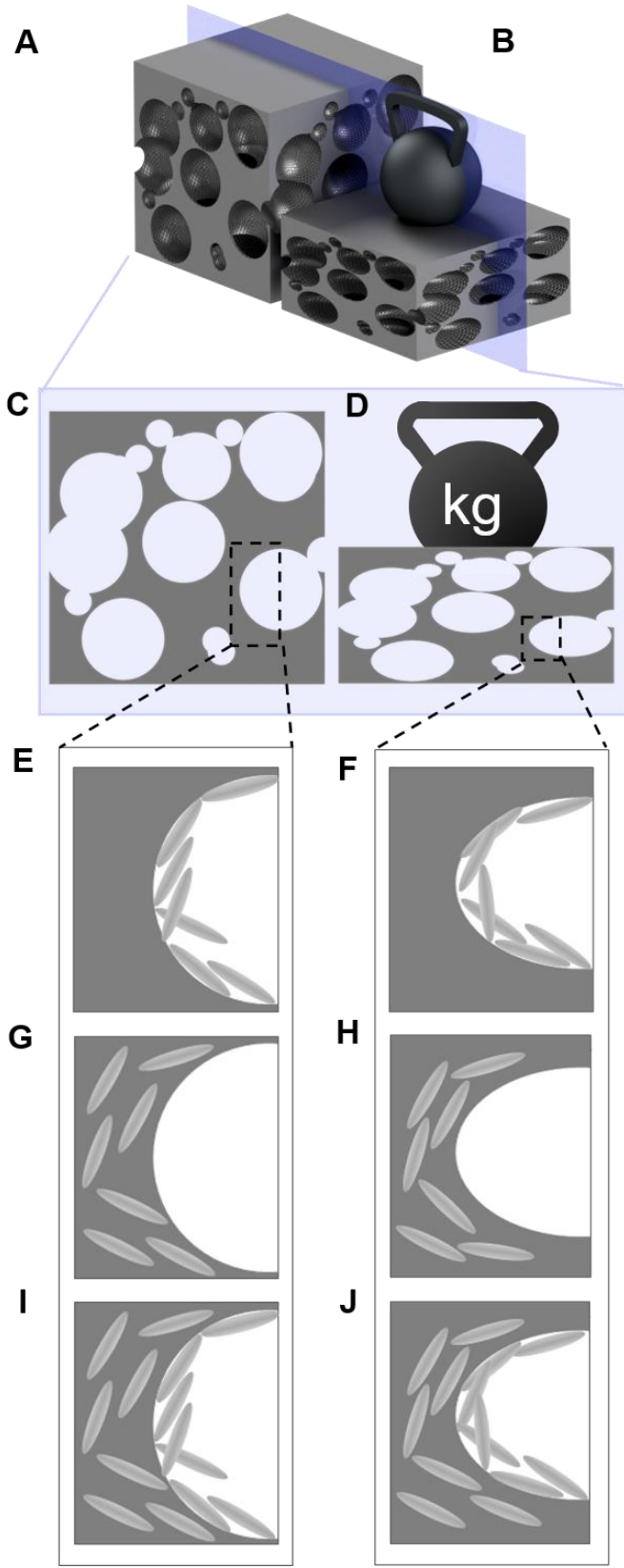

**Figure 4**. Schematic illustration of changes in the GPFs upon application of pressure. **(A, B)** Three dimensional and **(C, D)** cross sectional view of GPFs when compressed illustrating the closing of pores. Distribution of graphene platelets when the foams are in uncompressed and compressed states for **(E, F)** GCFs; **(G, H)** GEFs; and **(I, J)** GCEFs.

In GCEFs, the conductive pathways occur both due to the formation of percolation networks as well as tunnelling effect, which are explained above. The concentration of graphene is much higher in GCEFs (compared to GCF and GEF) which may enhance the conductivity. When pressure is applied, both the percolation network formation as well as the increase in tunnelling effect happens due to the presence of coated as well as incorporated graphene flakes (Figure 4I & J).

### 2.4. Pressure Sensor

The GPFs were incorporated in a simple device structure, shown in **Figure 5**A, to form a capacitive pressure sensor and is the sensing unit. The top carbon electrode is in direct contact with the GPF, thus forming a foam electrode, which has the ability to transfer charges effectively.[37] The bottom electrode is insulated with a very thin layer of PET (2 μm) to limit the fringing effect, ultimately reducing the power dissipation.[16,18] Inset of Figure 5A is a photograph of the pressure sensor enclosed between electrical insulating tapes. A capacitive pressure sensor works on the principle of a parallel plate capacitor and capacitance, C, is given by Equation 2.

$$C = \varepsilon_o \varepsilon_r \frac{A}{d} \qquad 2$$

where $\varepsilon_o$ is the permittivity of free space, $\varepsilon_r$ is the permittivity of dielectric, A is the overlapping area between the two electrodes, and d is the distance between the two electrodes. Upon application of pressure, the distance d reduces due to the geometrical deformation of the foam, which results in an increase in capacitance according to the Equation 2. The dielectric property of the foam also changes as air pockets are replaced with the solid PDMS upon application of pressure and contributes to the change in capacitance.

Figure 5B represents the average capacitances of the pressure sensors when no pressure is applied. Capacitances for the 2 mm foams are higher than the 4 mm foams, which is expected following equation 2, but does not exactly satisfy equation 2, i.e., the capacitance for 2 mm foams are not double the capacitance of 4 mm foams (Table S2). Such deviation from equation 2 has been reported in previous articles.[30] For all types of GPFs, the initial capacitances are not significantly different (Table S2), which implies that the way graphene is incorporated in the GPFs does not make a significant difference in $\varepsilon_r$ of the foams. However, when compared to PDMS only foams, GPFs have higher initial capacitance (Table S2, S3). This suggest that the addition of graphene to PDMS increases the $\varepsilon_r$ of the foams.

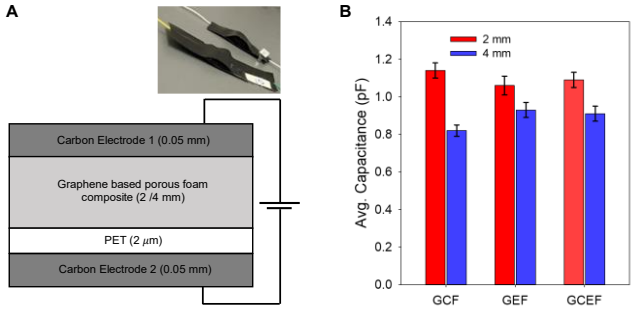

**Figure 5.** **(A)** Schematic of the structure of capacitive pressure sensor. Inset: photograph of a pressure sensor. **(B)** Average capacitance of pressure sensors when no pressure is applied. Error bars denote one standard deviation.

The responses of the pressure sensors with respect to increase in applied pressure are shown in **Figure 6**A-C. Error bars denote one standard deviation. The plots clearly indicate a monotonic increase in capacitance with increase in applied pressure. However, from the relative change in capacitance plots (Figure 6D-F), there are two distinct linear regions with different slopes: region A spanning roughly 0 – 6 kPa and region B spanning roughly 6 – 12 kPa. These regions correspond to region I and region II of the stress-strain curves (Figure 3). For all types of foams, where there is higher strain to applied stress for 0 – 6 kPa. In region A (Figures 6D-F), there are two factors contributing to change in capacitance. With applied pressure, the low dielectric air pores in the foams are getting compressed and replaced by the higher dielectric solid of PDMS + graphene. The applied pressure will also reduce the thickness of the foam. Therefore, change in dielectric constant and thickness of the foam both contributes to the increase in capacitance. As pressure increases, the air pores are completely replaced with PDMS + graphene and region B starts at this point. Further increase in pressure results only in compression of the PDMS + graphene. Hence, there is no change in the dielectric constant of the foam. Therefore, the change in capacitance in region II is coming only from the change in thickness due to the applied pressure. Also, the relative change in thickness of foam in region B will be lesser than region A as there are no air pores in the foam when it reaches region B.[25] Therefore, the relative change in capacitance is lesser for region B as compared to region A.

In terms of the sensitivities of the pressure sensors, the presence of graphene in GPFs makes a significant difference as compared to PDMS-only foam. The sensitivities for all types of GPF pressure sensors are ~ 3 times higher than pressure sensors with PDMS-only foam (Table S2, S3). For all types of foams, the sensitivities are higher for 2 mm thick pressure sensors (Figure 6D-F). This is because the strain % is higher for the 2 mm foams for all three types of foams (Figures 3A – C, Table S1). The sensitivities for the pressure sensors also follow the strain % relationship.[25] The GCF has the highest strain % and hence the highest sensitivities followed by GEFs and GCEFs. The only anomaly in this pattern is the sensitivity for 4 mm GCEFs. As expected, the sensitivities are higher for region A as compared to region B for all the GPF pressure sensors since relative change in capacitance is higher in region A.

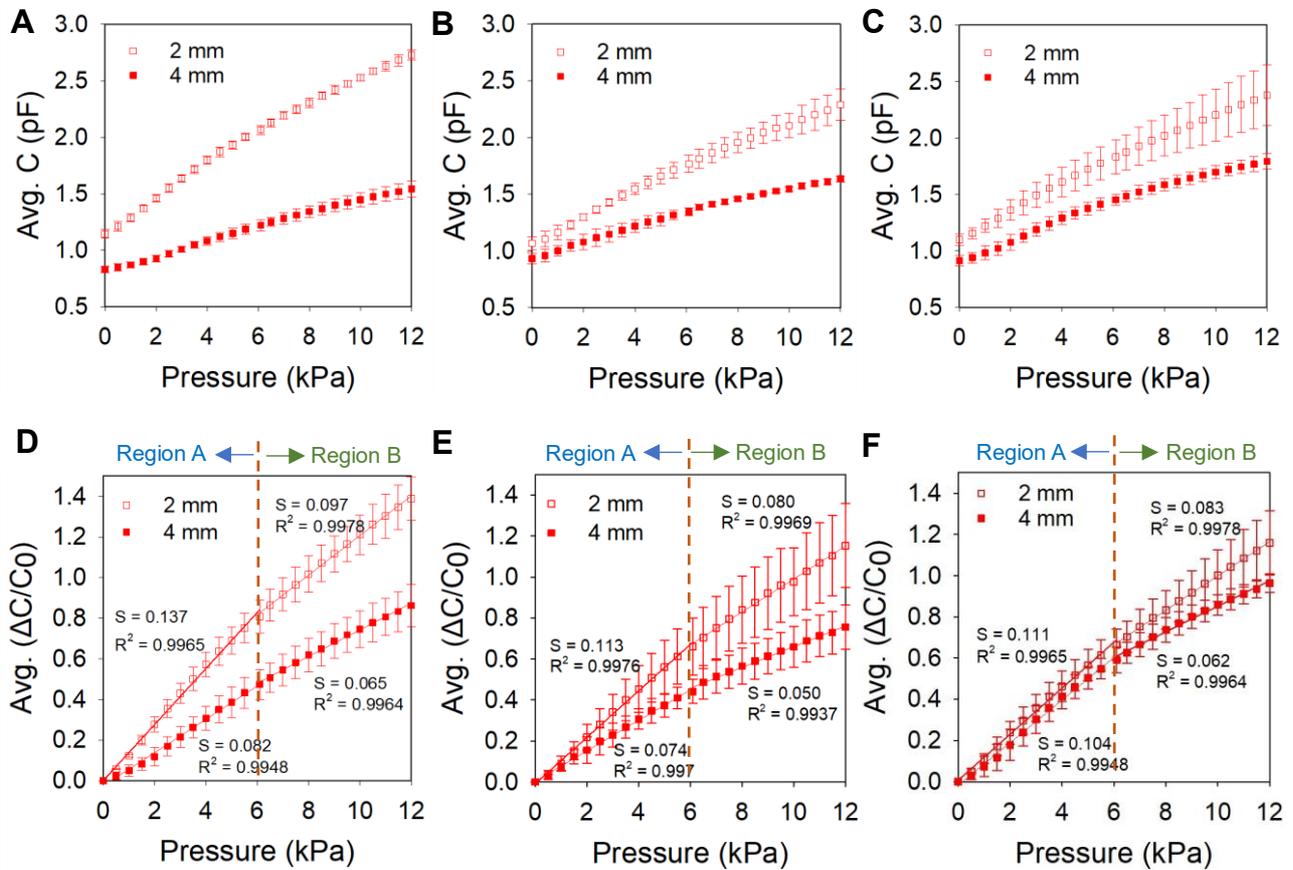

**Figure 6.** Average capacitance (C) variation with pressure for the three sensors with **(A)** GCF; **(B)** GEF; and **(C)** GCEF as the sensing units. Average relative change in capacitance with respect to initial capacitance (no applied pressure) for **(D)** GCF; **(E)** GEF; and **(F)** GCEF pressure sensors.

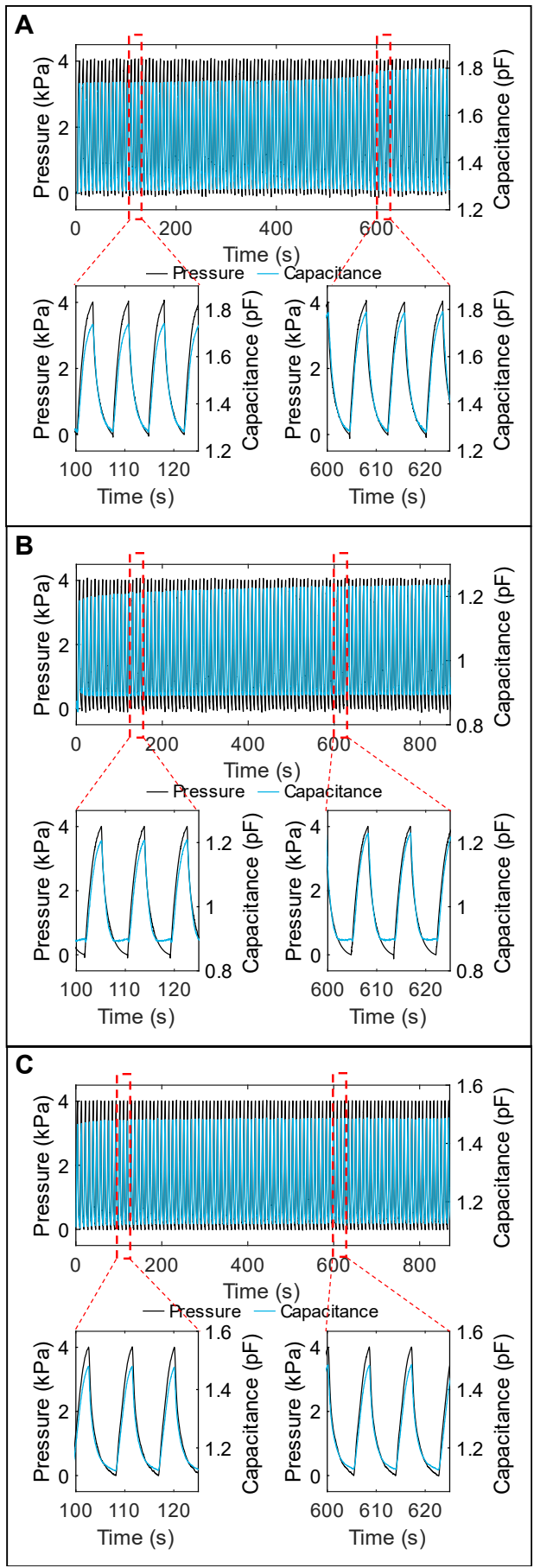

**Figure 7**. Durability of the pressure sensors were tested by subjecting the pressure sensors to 100 cycles of pressure at 4 kPa. Response of pressure sensors with 2 mm thick **(A)** GCF **(B)** GEF and **(C)** GCEF.

The limits of detection (LoD) for the pressure sensors are 0.14 (0.49), 0.32 (1.2) and 0.28 (0.8) for 2 mm (4 mm) GCF, GEF and GCEF, respectively (Table S4). These LoDs are one of the lowest reported in literature.[38] LoDs for GPFs which has graphene coated in the pores (GCF and GCEF) are lower than graphene embedded in the foam (GEF) because when pressure is applied there is higher conductivity in the foam due to graphene flakes coming in close proximity to each other as these have a greater freedom to move on the surface of the pore (Figure 4). Further, the GCF has a lower LoD than the GCEF because the GCEF is stiffer due to the embedded graphene in the PDMS matrix.

The durability of the pressor sensors was tested by subjecting them to 100 loading-unloading cycles at a constant pressure of 4 kPa. **Figure 7** shows the response of the pressure sensors with 2 mm thickness for all types of GPFs for the durability test. There is very little to no significant change in the response of the pressure sensors after 100 cycles, which shows that these pressure sensors can withstand multiple loading and unloading cycles required in most practical applications. In Figure 7, the time taken to complete 100 loading and unloading cycles are different for the different GPF pressure sensors and is reflective of the difference in stiffness of the GPFs. GCF pressure sensor took the least time to complete the 100 cycles, followed by GEF pressure sensors and GCEF pressure sensors. GCF has the least stiffness since graphene is present only as coatings of the pores. Stiffness is more for GEF than GCF due to the presence of graphene inside the foam. GCEF has the highest stiffness because of the higher concentration of graphene present within and outside the PDMS matrix. Response of durability test for pressure sensors with 4 mm thick foams are shown in Figure S7. Similar to the response of pressure sensors with 2mm foams, the relative capacitance remains the same for all the sensors over the 100 cycles.

## 2.5. Applications

Finally, multiple potential applications were demonstrated to show the versatility of the GPF pressure sensors. **Figure 8**A shows that the pressure sensor can easily pick up signals from light finger tapping and accurately determine the number of times of finger taps (one to five). The index finger was used for tapping on the pressure sensor consecutively with small breaks. Figure 8B shows detection of metacarpal bone movement by placing the pressure sensor on top of the metacarpal region. The folding of hand corresponds to increase in capacitance and the unfolding of hand corresponds to decrease in capacitance. The pressure sensors can also detect the respiratory rate when placed on the inside of a waist belt. Figure 8C shows difference in pressure sensor response corresponding to relatively slow and fast respiration rates. Lastly, the pressure sensor can pick up signals from swallowing movements when placed on the throat. It is also able to detect the difference in quantity of water swallowed as shown in Figure 8D. These demonstrations show the feasibility of using our pressure sensor for various applications.

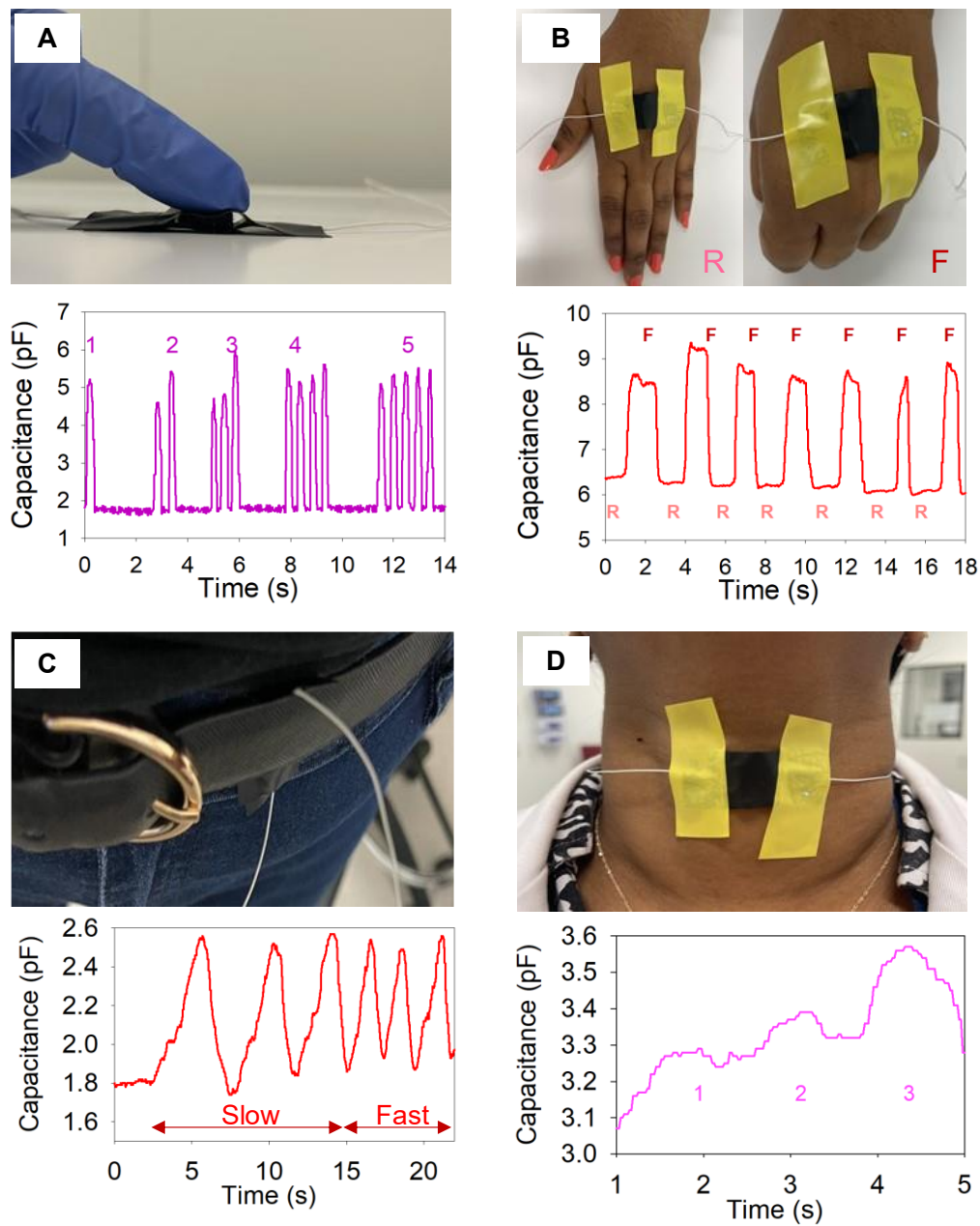

**Figure 8.** Applications of pressure sensor. Photographs of application and pressure sensor response for **(A)** detection of finger tapping with the number of taps; **(B)** detection of metal carpal movements during folding of hand into fist and relaxed state; **(C)** detection of respiratory rate by putting pressure sensor below the belt; and **(D)** Detection of swallowing movement for different quantities of water (1 < 2 < 3).

## 3. Conclusion

In conclusion, we fabricated GPFs using simple sugar templating method. Graphene was integrated into the foam in different ways: as coatings of the pores of the foam (GCF), embedded in the PDMS matrix of the foam (GEF) and combination of both (GCEF). The GPFs obtained are highly porous (~ 80% porosity), lightweight, squeezable as well as bendable. Capacitive pressure sensors fabricated using the GPFs can efficiently detect low pressures < 12 kPa. Sensitivities are higher sensitivity for pressures < 6 kPa. The pressure sensors have very low LoD, in the order of magnitude $1.0 \times 10^{-1}$ Pa. The difference in performance of pressure sensors arises due to difference in stress-strain relationships as well as the difference internal mechanism of the GPFs. The GPFs pressure sensors are very stable and shows no significant change upon repeated testing (100 cycles). Lastly, the potential for real world applications of the pressure sensors developed were demonstrated through detection of swallowing motion, respiration rate, etc. Further optimization of pressure sensor parameters such as the amount of graphene in the foam, pore sizes, mechanical strength of foam, etc. could further enhance the performance of the pressure sensors in future. Our work shows the potential of graphene porous foam based capacitive pressure sensors for detection of low pressures, paving the way for further research in this area.

## 4. Experimental Section

First, to obtain templates of the foams, commercially available sugar cubes of sizes 20 mm × 20 mm × 20 mm were sanded to reduce thickness to 2 mm and 4 mm (see Figure S1).

**GCF:** To make PDMS foams, sugar templates were dropped in PDMS precursor (RTV Momentive 615) mixed with curing agent (0.55 g) at a ratio of 10:1. This was transferred into a vacuum chamber for 15 minutes at $1\times10^{-1}$ torr to remove the air bubbles and accelerate the capillary action of infiltrating sugar templates with PDMS mixture. After this, the PDMS loaded sugar cube was left for curing in an oven at a temperature of 100°C for 30 minutes. Then, the cured PDMS loaded sugar cube was kept in water for 24 hours to dissolve the sugar skeleton leaving behind only the PDMS as a porous foam. The PDMS foam was dried by placing it in the oven for 30 minutes at 100°C. To obtain GCF, the PDMS was dipped in a solution of graphene nanoplatelets (Sigma-Aldrich) in isopropanol (IPA). The graphene solution was prepared by mixing 2 wt.% of graphene nanoplatelets with respect to mass ratio of PDMS (1.1 mg), in 3 mL of isopropanol. For homogenous mixing of graphene flakes, the solution was tip-sonicated for 30 minutes. The PDMS foam was kept submerged in the graphene for 10 minutes, followed by drying in the oven for 15 minutes at 80°C (see Figure S2).

**GEF:** PDMS precursor (0.55g) was diluted in 3 mL of cyclohexane (Sigma-Aldrich) and meanwhile a 1.1mg (to make 2% wt) of graphene nano-platelets was sonicated in 3 mL of IPA to form the graphene-PDMS solution. The curing agent was added into the mixture followed by the dropping of the sugar cube. This was kept in a vacuum chamber for 15 minutes at $1\times10^{-1}$ torr to remove the air bubbles and accelerate the capillary action of graphene-PDMS solution. After this, the graphene-PDMS loaded sugar cube was cured in an oven at a temperature of 100°C for 30 minutes. The cured graphene-PDMS sugar cube was left in water for 24 hours for the dissolution of the sugar skeleton, followed by heating in an oven for 30 minutes at 100°C (see Figure S3).

**GCEF:** This was obtained by dipping GEF in graphene solution (1.1mg in 3mL of IPA) to form a GCEF.

**Pressure sensors:** All porous foams were cut to 10 mm × 10 mm squares. Pressure sensors were fabricated by sandwiching GPFs between two flexible carbon tapes of thickness 0.05 mm (ProSciTech) that act as electrodes (Figure 5). Insulating PET layer is inserted between one electrode and the GPF. The pressure sensors were enclosed in insulating tapes.

**Characterization:** The mass of the GPFs was measured by electronic precision balance (Kern & Shon). SEM images of the GPFs were obtained using a SEM - PHENOM XL G2 under different magnifications. The GPF samples were coated with 2-5 nm of Pt to improve the imaging. Stress-strain relationship of the foams were obtained by uniaxial compression tests conducted by increasing the stress up to 20 kPa (Mecmesin- Multi-test 2.5-dv). Capacitance measurements of the pressure sensors were obtained using a benchtop LCR meter (BK Precision 891) set at frequency of 1 kHz and a voltage of 1 Volt. The pressures on the sensors were applied using a high precision universal testing machine (Mecmesin- Multi-test 2.5-dv) which applied normal force on the sensors. Five samples for each type of foam were tested to get the averages.

The measurements on human body were done in accordance with the guidelines of University Human Research Ethics Committee at Queensland University of Technology (QUT) (ethics approval number: 2021000254).

## Author Contributions
L.K. fabricated, characterised, and analysed data. L.K. and CC imaged the foams. J.A. C.C. and L.K. established fabrication techniques and characterization set-ups. S.Y. developed the idea and supervised the project. All authors contributed to drafting, reviewing and editing the manuscript.

## Conflicts of interest
There are no conflicts to declare.

## Acknowledgements
This work was partly carried out at the Central Analytical Research Facility (CARF) at QUT. We would like to thank CARF staffs and technicians for their support and help in using various equipment. L.K.'s work was supported through an Australian Government Research Training Program Scholarship.

## References

1. Luo, Z.; Chen, J.; Zhu, Z.; Li, L.; Su, Y.; Tang, W.; Omisore, O. M.; Wang, L.; Li, H. High-Resolution and High-Sensitivity Flexible Capacitive Pressure Sensors Enhanced by a Transferable Electrode Array and a Micropillar–PVDF Film. *ACS Appl. Mater. Interfaces* **2021**, *13*, 7635-7649.

2. Pang, Y.; Zhang, K.; Yang, Z.; Jiang, S.; Ju, Z.; Li, Y.; Wang, X.; Wang, D.; Jian, M.; Zhang, Y.; Liang, R.; Tian, H.; Yang, Y.; Ren, T.L. Epidermis Microstructure Inspired Graphene Pressure Sensor with Random Distributed Spinosum for High Sensitivity and Large Linearity. *ACS Nano* **2018**, *12*, 2346-2354.

3. Zeng, X.; Wang, Z.; Zhang, H.; Yang, W.; Xiang, L.; Zhao, Z.; Peng, L. M.; Hu, Y. Tunable, Ultrasensitive, and Flexible Pressure Sensors Based on Wrinkled Microstructures for Electronic Skins. *ACS Appl. Mater. Interfaces* **2019**, *11*, 21218-21226.

4. Boutry, C. M.; Beker, L.; Kaizawa, Y.; Vassos, C.; Tran, H.; Hinckley, A. C.; Pfattner, R.; Niu, S.; Li, J.; Claverie, J.; Wang, Z.; Chang, J.; Fox, P. M.; Bao, Z. Biodegradable and flexible arterial-pulse sensor for the wireless monitoring of blood flow. *Nat. Biomed. Eng.* **2019**, *3*, 47.

5. Wang, J.; Suzuki, R.; Shao, M.; Gillot, F.; Shiratori, S. Capacitive Pressure Sensor with Wide-Range, Bendable, and High Sensitivity Based on the Bionic Komochi Konbu Structure and Cu/Ni Nanofiber Network. *ACS Appl. Mater. Interfaces* **2019**, *11*, 11928-11935.

6. Kim, S. j.; Phung, T. H.; Kim, S.; Rahman, M. K.; Kwon, K. S. Low-Cost Fabrication Method for Thin, Flexible, and Transparent Touch Screen Sensors. *Adv. Mater. Technol.* **2020**, *5*, 2000441.

7. Wu, C.C. Ultra-high transparent sandwich structure with a silicon dioxide passivation layer prepared on a colorless polyimide substrate for a flexible capacitive touch screen panel. *Sol. Energy Mater. Sol.* **2020**, *207*, 110350.

8. Li, J.; Bao, R.; Tao, J.; Peng, Y.; Pan, C. Recent progress in flexible pressure sensor arrays: from design to applications. *J. Mater. Chem. C* **2018**, *6*, 11878-11892.

9. Li, W.; Xiong, L.; Pu, Y.; Quan, Y.; Li, S. High-Performance Paper-Based Capacitive Flexible Pressure Sensor and Its Application in Human-Related Measurement. *Nanoscale. Res. Lett.* **2019**, *14*, 1-7.

10. Chang, Y.; Zuo, J.; Zhang, H.; Duan, X. State-of-the-art and recent developments in micro/nanoscale pressure sensors for smart wearable devices and health monitoring systems. *Nanotechnology and Precision Engineering* **2020**, *3*, 43-52.

11. Nag, M.; Singh, J.; Kumar, A.; Alvi, P. A.; Singh, K. Sensitivity enhancement and temperature compatibility of graphene piezoresistive MEMS pressure sensor. *Microsyst. Technol.* **2019**, *25*, 3977-3982.

12. Meng, Z.; Zhang, H.; Zhu, M.; Wei, X.; Cao, J.; Murtaza, I.; Ali, M. U.; Meng, H.; Xu, J. Lead Zirconate Titanate (a piezoelectric ceramic)-Based thermal and tactile bimodal organic transistor sensors. *Organ. Electron.* **2020**, *80*, 105673.

13. Guo, W.; Tan, C.; Shi, K.; Li, J.; Wang, X. X.; Sun, B.; Huang, X.; Long, Y.-Z.; Jiang, P. Wireless piezoelectric devices based on electrospun PVDF/BaTiO3 NW nanocomposite fibers for human motion monitoring. *Nanoscale* **2018**, *10*, 17751-17760.

14. Shi, Q.; Wang, H.; Wang, T.; Lee, C. Self-powered liquid triboelectric microfluidic sensor for pressure sensing and finger motion monitoring applications. *Nano Energy* **2016**, *30*, 450-459.

15. Yang, D.; Guo, H.; Chen, X.; Wang, L.; Jiang, P.; Zhang, W.; Zhang, L.; Wang, Z. L. A flexible and wide pressure range triboelectric sensor array for real-time pressure detection and distribution mapping. *J. Mater. Chem. A* **2020**, *8*, 23827-23833.

16. Pruvost, M.; Smit, W. J.; Monteux, C.; Poulin, P.; Colin, A. Polymeric foams for flexible and highly sensitive low-pressure capacitive sensors. *npj Flex. Electron.* **2019**, *3*, 7.

17. Park, S. W.; Das, P. S.; Chhetry, A.; Park, J. Y. A Flexible Capacitive Pressure Sensor for Wearable Respiration Monitoring System. *IEEE Sens. J.* **2017**, *17*, 6558-6564.

18. Pruvost, M.; Smit, W. J.; Monteux, C.; Poulin, P.; Colin, A. Microporous electrostrictive materials for vibrational energy harvesting. *Multifunctional Materials* **2018**, *1*, 015004.

19. Madsen, F. B.; Daugaard, A. E.; Hvilsted, S.; Skov, A. L. The Current State of Silicone-Based Dielectric Elastomer Transducers. *Macromol. Rapid Commun.* **2016**, *37*, 378-413.

20. Mannsfeld, S. C. B.; Tee, B. C. K.; Stoltenberg, R. M.; Chen, C. V. H. H.; Barman, S.; Muir, B. V. O.; Sokolov, A. N.; Reese, C.; Bao, Z. Highly sensitive flexible pressure sensors with microstructured rubber dielectric layers. *Nat. Mater.* **2010**, *9*, 859-864.

21. Park, J.; Lee, Y.; Hong, J.; Ha, M.; Jung, Y. D.; Lim, H.; Kim, S. Y.; Ko, H. Giant Tunneling Piezoresistance of Composite Elastomers with Interlocked Microdome Arrays for Ultrasensitive and Multimodal Electronic Skins. *ACS Nano* **2014**, *8*, 4689-4697.

22. Chen, Y. M.; He, S. M.; Huang, C. H.; Huang, C. C.; Shih, W. P.; Chu, C. L.; Kong, J.; Li, J.; Su, C.-Y. J. N. Ultra-large, suspended graphene as a highly elastic membrane for capacitive pressure sensors. *Nanoscale* **2016**, *8*, 3555-3564.

23. Chen, X.; Li, X.; Shao, J.; An, N.; Tian, H.; Wang, C.; Han, T.; Wang, L.; Lu, B. High-Performance Piezoelectric Nanogenerators with Imprinted P(VDF-TrFE)/BaTiO3 Nanocomposite Micropillars for Self-Powered Flexible Sensors. *Small* **2017**, *13*, 1604245.

24. Atalay, O.; Atalay, A.; Gafford, J.; Walsh, C. A Highly Sensitive Capacitive-Based Soft Pressure Sensor Based on a Conductive Fabric and a Microporous Dielectric Layer. *Adv. Mater. Technol.* **2018**, *3*, 1700237.

25. Choi, J.; Kwon, D.; Kim, K.; Park, J.; Orbe, D. D.; Gu, J.; Ahn, J.; Cho, I.; Jeong, Y.; Oh, Y.; Park, I. Synergetic Effect of Porous Elastomer and Percolation of Carbon Nanotube Filler toward High Performance Capacitive Pressure Sensors. *ACS Appl. Mater. Interfaces* **2020**, *12*, 1698-1706.

26. Jeon, D. Y.; Kim, H.; Lee, M. W.; Park, S. J.; Kim, G. T. Piezo-impedance response of carbon nanotube/polydimethylsiloxane nanocomposites. *APL Mater.* **2019**, *7*, 041118.

27. Cataldi, P.; Athanassiou, A.; Bayer, I. S., Graphene Nanoplatelets-Based Advanced Materials and Recent Progress in Sustainable Applications. *J. Appl. Sci.* **2018**, *8*, 1438.

28. Ke, K.; McMaster, M.; Christopherson, W.; Singer, K. D.; Manas-Zloczower, I. Highly sensitive capacitive pressure sensors based on elastomer composites with carbon filler hybrids. *Compos. Part A Appl. Sci. Manuf.* **2019**, *126*, 105614.

29. Jung, Y.; Jung, K.; Park, B.; Choi, J.; Kim, D.; Park, J.; Ko, J.; Cho, H. Wearable piezoresistive strain sensor based on graphene-coated three-dimensional micro-porous PDMS sponge. *Micro Nano Syst. Lett.* **2019**, *7*, 20.

30. Bilent, S.; Dinh, T. H. N.; Martincic, E.; Joubert, P. Y. Influence of the Porosity of Polymer Foams on the Performances of Capacitive Flexible Pressure Sensors. *Sensors* **2019**, *19*, 1968.

31. Boland, C. S.; Khan, U.; Backes, C.; O'Neill, A.; McCauley, J.; Duane, S.; Shanker, R.; Liu, Y.; Jurewicz, I.; Dalton, A. B.; Coleman, J. N. Sensitive, High-Strain, High-Rate Bodily Motion Sensors Based on Graphene–Rubber Composites. *ACS Nano* **2014**, *8*, 8819-8830.

32. Sengupta, D.; Pei, Y.; Kottapalli, A. G. P. Ultralightweight and 3D squeezable graphene-polydimethylsiloxane composite foams as piezoresistive sensors. *ACS Appl. Mater. Interfaces* **2019**, *11*, 35201-35211.

33. Zhai, W.; Xia, Q.; Zhou, K.; Yue, X.; Ren, M.; Zheng, G.; Dai, K.; Liu, C.; Shen, C. Multifunctional flexible carbon black/polydimethylsiloxane piezoresistive sensor with ultrahigh linear range, excellent durability and oil/water separation capability. *Chem. Eng. J.* **2019**, *372*, 373-382.

34. Oskouyi, A. B.; Sundararaj, U.; Mertiny, P. J. M. Tunneling conductivity and piezoresistivity of composites containing randomly dispersed conductive nano-platelets. *Materials* **2014**, *7*, 2501-2521.

35. Hicks, J.; Behnam, A.; Ural, A. A computational study of tunneling-percolation electrical transport in graphene-based nanocomposites. *Appl. Phys. Lett.* **2009**, *95*, 213103.

36. Hempel, M.; Nezich, D.; Kong, J.; Hofmann, M. A novel class of strain gauges based on layered percolative films of 2D materials. *Nano Lett.* **2012**, *12*, 5714-5718.

37. Liu, J.; Zhang, L.; Wu, H. B.; Lin, J.; Shen, Z.; Lou, X. W. High-performance flexible asymmetric supercapacitors based on a new graphene foam/carbon nanotube hybrid film. *Energy & Environ. Sci.* **2014**, *7*, 3709-3719.

38. Cicek, M. O.; Doganay, D.; Durukan, M. B.; Gorur, M. C.; Unalan, H. E. Seamless Monolithic Design for Foam Based, Flexible, Parallel Plate Capacitive Sensors. *Adv. Mater. Technol.* **2021**, *6*, 2001168.

# Graphene porous foams for capacitive pressure sensing

*Lekshmi A. Kurup[†], Cameron M. Cole[†], Joshua N. Arthur[‡], and Soniya D. Yambem[†] *

[†]School of Chemistry and Physics, Centre for Materials Science, Faculty of Science, Queensland University of Technology (QUT), 2 George Street, Brisbane, QLD 4000, Australia.

*Email: soniya.yambem@qut.edu.au

Commercially bought sugar cubes are resized manually into 2 mm and 4 mm thick sugar cubes using sandpaper and the thickness is confirmed by means of a vernier calipers.

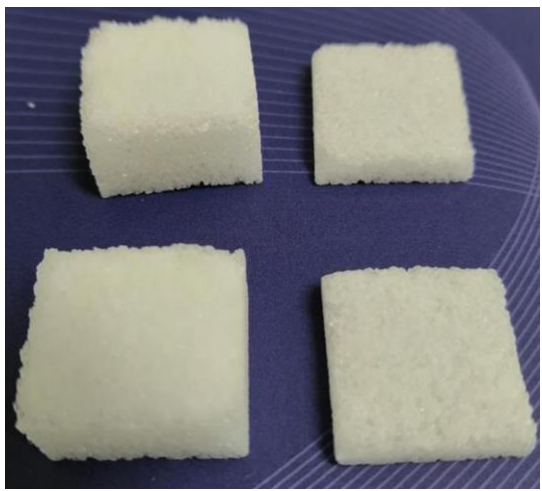

**Figure S1.** Commercially available sugar cubes (left) and resized sugar cubes (right).

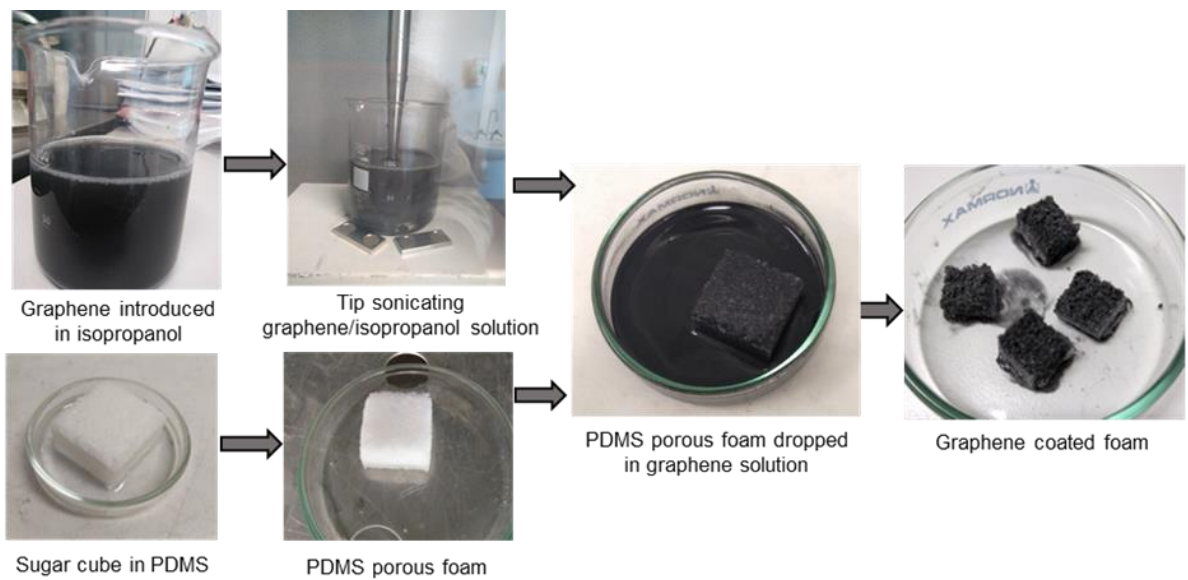

**Figure S2.** Fabrication of graphene coated foams (GCF).

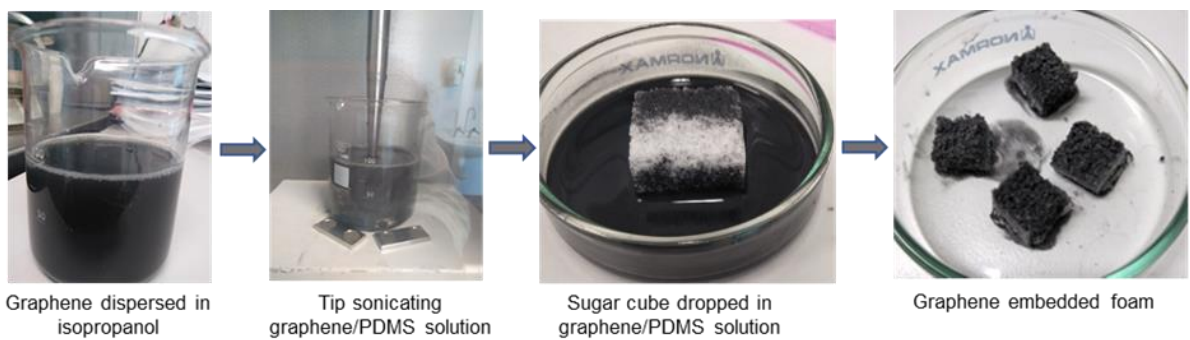

**Figure S3.** Fabrication of graphene embedded porous (GEF).

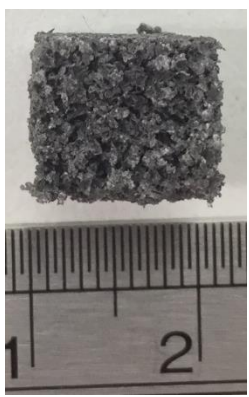

**Figure S4.** Photograph of a graphene coated foam under a magnified lens.

**Porosity calculation**

For these calculations, the mass of PDMS were the measured mass of the foam, which also includes graphene. Since, the weight of graphene is negligible, we have taken the mass of foam as equivalent to mass of PDMS.

*Porosity of 2 mm foams*

Porosity = ratio of volume of air pores to the volume of the foam composite.

Volume of the foam composite = 1 x 1 x 0.2 = 0.2 cm$^2$

Volume of air pores = Volume of foam composite – Volume of PDMS

Volume of PDMS = $\frac{Mass\ of\ PDMS\ (g)}{Density\ of\ PDMS\ (g/cm^3)} = \frac{0.039}{0.9625} = 0.040$

∴ Volume of air pores = 0.2 – 0.040 = 0.16

Porosity (%) = $\frac{0.16}{0.2}$ = 80%

*Porosity of 4 mm foams*

Porosity = ratio of volume of air pores to the volume of the foam composite.

Volume of the foam composite = 1 x 1 x 0.4 = 0.4 cm$^2$

Volume of air pores = Volume of foam composite – Volume of PDMS

Volume of PDMS = $\frac{Mass\ of\ PDMS\ (g)}{Density\ of\ PDMS\ (g/cm^3)} = \frac{0.077}{0.9625} = 0.080$

∴ Volume of air pores = 0.4 – 0.080 = 0.32

Porosity (%) = $\frac{0.32}{0.4}$ = 80%

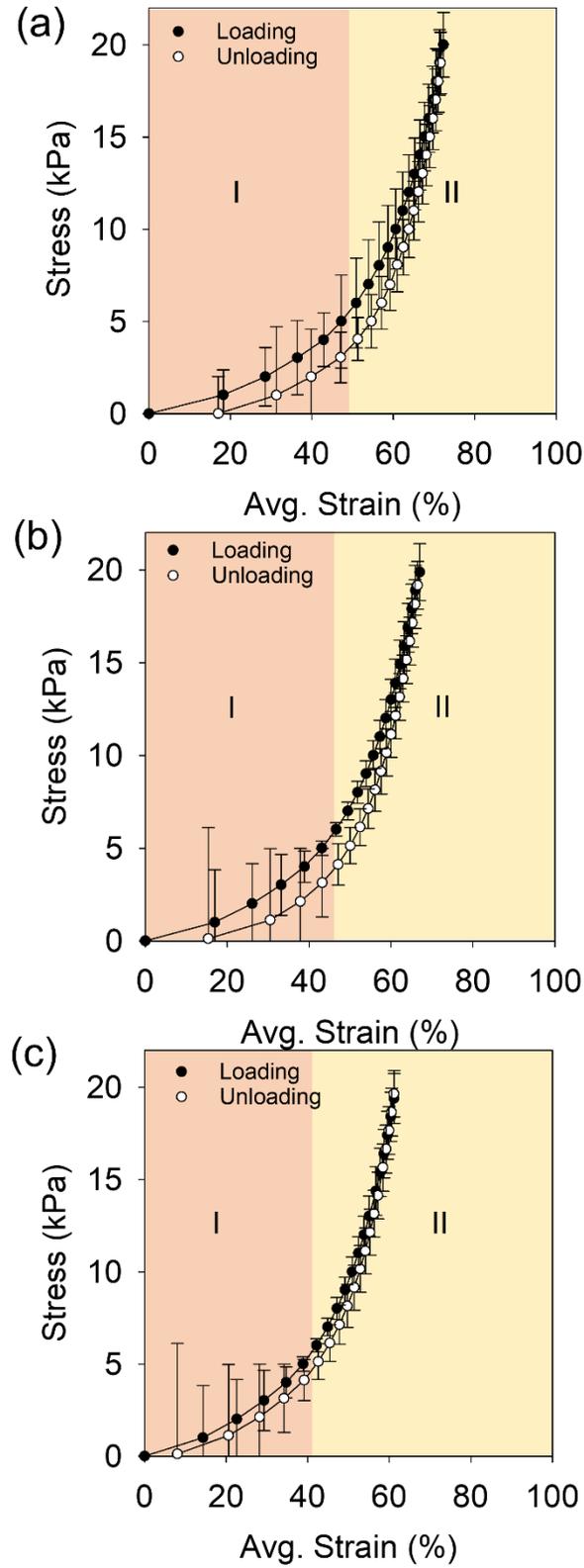

**Figure S5.** Stress-Strain curves with a maximum stress of 20kPa for 4 mm thick (a) GCF; (b) GEF; and (c) GCEF.

**Table S1.** Maximum strain % at a stress of 20 kPa

| Thickness/Foam Type | GCF | GEF | GCEF |
|---|---|---|---|
| 2 mm | 82.7 ± 0.64 | 72.3 ± 0.75 | 66.7 ± 1.08 |
| 4 mm | 71.93 ± 0.61 | 66.9 ± 0.93 | 61.1 ± 1.26 |

**Table S2.** Average capacitances (initial = no pressure) and sensitivities of the pressure sensors for different types and thickness of graphene porous foams.

| Type of Foam composite | Thickness (mm) | Average Initial Capacitance (pF) | Average Sensitivity (kPa$^{-1}$) | |
|---|---|---|---|---|
| | | | Low Pressure (0-6 kPa) | High Pressure (6-12 kPa) |
| GCF | 2 | 1.14 ± 0.04 | 0.137 ± 0.002 | 0.097 ± 0.001 |
| | 4 | 0.82 ± 0.03 | 0.082 ± 0.001 | 0.065 ± 0.001 |
| GEF | 2 | 1.06 ± 0.05 | 0.113 ± 0.001 | 0.080 ± 0.001 |
| | 4 | 0.93 ± 0.04 | 0.074 ± 0.001 | 0.050 ± 0.001 |
| GCEF | 2 | 1.09 ± 0.04 | 0.111 ± 0.001 | 0.083 ± 0.000 |
| | 4 | 0.91 ± 0.04 | 0.104 ± 0.002 | 0.062 ± 0.001 |

**Table S3.** Average capacitances (initial = no pressure) and sensitivities of the pressure sensor with PDMS-only porous foam.

| Type of Foam composite | Thickness (mm) | Average Initial Capacitance (pF) | Average Sensitivity (kPa$^{-1}$) | |
|---|---|---|---|---|
| | | | Low Pressure (0-6 kPa) | High Pressure (6-12 kPa) |
| PDMS | 2 | 0.97 ± 0.09 | 0.0402 ± 0.000 | 0.0321 ± 0.001 |
| | 4 | 0.65 ± 0.01 | 0.030 ± 0.000 | 0.0251 ± 0.000 |

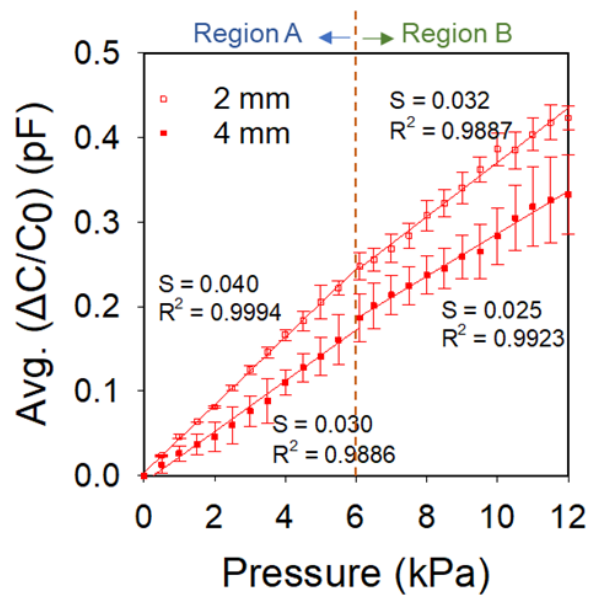

**Figure S6.** Relative change in capacitance of pressure sensors with a PDMS-only porous foam.

**Table S4.** Average limit of detection for pressure sensor with different types of foams.

| Type of Foam | Thickness (mm) | Limit of Detection (LOD) (Pa) |
|---|---|---|
| GCF | 2 | 0.14 ± 0.004 |
|  | 4 | 0.49 ± 0.03 |
| GEF | 2 | 0.32 ± 0.01 |
|  | 4 | 1.2 ± 0.06 |
| GCEF | 2 | 0.28 ± 0.008 |
|  | 4 | 0.82 ± 0.05 |
| PDMS | 2 | 0.37 ± 0.03 |
|  | 4 | 1.5 ± 0.301 |

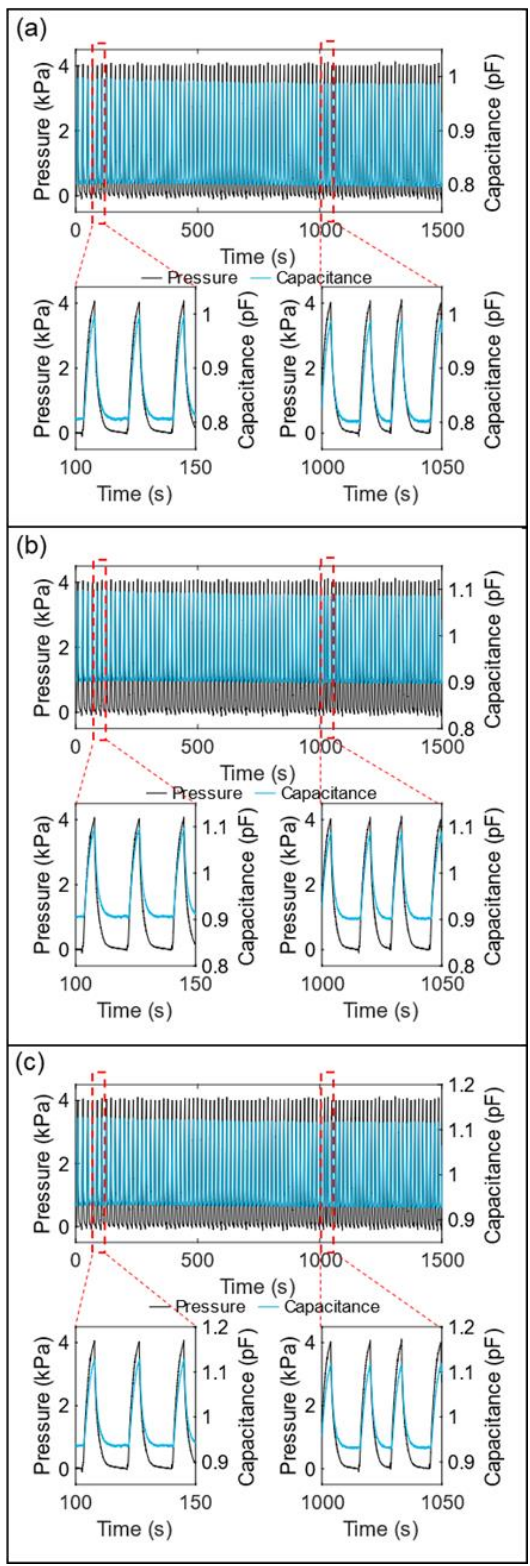

**Figure S7**. Durability test for 4mm thick GPF pressure sensors. Response of pressure with (a) GCF; (b) GEF; and (c) GCEF for 100 cycles of maximum pressure 4 kPa.